# Disperse rotation operator DRT and use in some stream ciphers

Yong-Jin Kim[1], Yong-Ho Yon[2], Son-Gyong Kim[3]

[1] Faculty of Mathematics, KIM IL SUNG University, Pyongyang, D. P. R of Korea
[2] Institute of Mathematics, National Academy of Sciences, Pyongyang, D. P. R of Korea
[3] Institute of Management Practice, Ministry of Information Industry, Pyongyang, D. P. R of Korea

Corresponding author: Yong-Jin Kim (kyj0916@126.com)

**Abstract:** The rotation operator is frequently used in several stream ciphers, including HC-128, Rabbit, and Salsa20, the final candidates for eSTREAM. This is because the rotation operator (ROT) is simple but has very good dispersibility.
In this paper, we propose a disperse rotation operator (DRT), which has the same structure as ROT but has better dispersibility. In addition, the use of DRT instead of ROT has shown that the quality of the output stream of all three stream ciphers was significantly improved. However, the use of DRT instead of ROT in the HC-128 stream cipher prevents the expansion of differential attacks based on LSB.

## 1. Introduction

The purpose of an update or output function of stream ciphers is to provide randomness and dispersion on the state array and output streams. Therefore, it is necessary for this function to achieve a better dispersion. Many stream ciphers, such as HC-128, Rabbit, and Salsa20, the portfolios of eSTREAM, often use the rotation operator ROT. The following equation was used in the keystream generation algorithm of HC-128 in [1].

$$P[j] = P[j] + g1(P[j \boxminus 3], P[j \boxminus 10], P[j \boxminus 511]) \qquad (1)$$

Here $g1(x, y, z) = (x \ggg 10) \oplus (z \ggg 23) + (y \ggg 8)$ and $\ggg, \lll$ are rotation operators. Similarly, the $Next\_State()$ function in [2] use the equation (2)

$$x_{0,i+1} = g_{0,i} + (g_{7,i} \lll 16) + (g_{6,i} \ggg 16) \qquad (2)$$

and the following one is used in $quarterround()$ function of [3].

$$z_1 = y_1 \oplus ((y_0 + y_3) \lll 7) \qquad (3)$$

Here $P, x, y, z, x_{0,i+1}, g_{0,i}, g_{7,i}, g_{6,i}, z_1, y_0, y_1$ and $y_3$ are the arrays of 32bit integers. As you can see, the ROT is used in all the above equations, and these equations form an algebraic structure to provide better dispersion on the internal state and output streams. To date, much research has been conducted on the more effective structure of the update/output function itself, but very little research has been done on the dispersibility of basic operators such as the ROT used in the functions.

In this paper, we presented a disperse rotation operator, DRT, which has the same structure as ROT, and provided the condition DRT to be a one-to-one function and considered its dispersibility. In addition, we show that the quality of the keystreams of HC-128, Rabbit, and Salsa20 will be improved remarkably when using DRT instead of ROT. However, if DRT is used instead of ROT in HC-128, then the result of [4] that the LSB-based distinguishing attack can also be expanded to other bits becomes impossible, so it will help in the security of HC-128.

# 2. DRT and its specification

Let the size of the operand be $n = 2^m$ bit, and then ROT mean left rotation operator $\lll$. The same is true for the right rotation operator $\ggg$. Subsequently, the ROT and DRT are defined as follows.

$$y = \text{ROT}(x, c) = (x \ll c) \oplus (x \gg (2^m - c)) \tag{4}$$

$$y = \text{DRT}(x, a, b) = (x \ll a) \oplus (x \gg b), a + b = 2^k < n \tag{5}$$

As can be seen, DRT has the same structure as ROT. The only difference is that ROT uses one parameter, whereas DRT uses two parameters.

The specific features of the DRT are as follows:

## 1.1 DRT is a one-to-one function.

Let us
$$a + b = 2^k < 2^m, \quad k = \overline{2, \ldots, m-1}, \quad x = (x_{2^m - 1} \cdots x_0)$$
and
$$y_1 = x \ll a = A_{2^{m-k+1}-1} || \ldots A_1 || A_0.$$
Here,

$$A_0 = (00 \cdots 0), \quad a \text{ bit}$$
$$A_{2i-1} = (x_{(i-1)*2^k + b - 1} \ x_{(i-1)*2^k + b - 2} \cdots x_{(i-1)*2^k}), \quad b \text{ bit}$$
$$A_{2i} = (x_{i*2^k - 1} \ x_{i*2^k - 2} \cdots x_{i*2^k - a}), \quad a \text{ bit}$$
$$i \in \{1, \ldots, 2^{m-k} - 1\}$$
$$A_{2^{m-k+1}-1} = (x_{2^m - 2^k + b - 1} \ x_{2^m - 2^k + b - 2} \cdots x_{2^m - 2^k}). \quad B \text{ bit}$$

The $||$ symbol is concatenation operator and the size of $A_0, A_2, \ldots, A_{2^{m-k+1}-2}$ is $a$ bit, otherwise, the size of $A_1, A_3, \ldots, A_{2^{m-k+1}-1}$ is $b$ bit. For example, if $m = 5$, $k = 4$, $a = 7$ and $b = 9$ then $x = (x_{31} \cdots x_0)$, $y_1 = x \ll 7 = A_3 \| A_2 \| A_1 \| A_0$. Here,

$$A_0 = (0000000),$$
$$A_1 = (x_8 \ x_7 \cdots x_0),$$
$$A_2 = (x_{15} \ x_{14} \cdots x_9),$$
$$A_3 = (x_{24} \ x_{23} \cdots x_{16}).$$

Similarly, let us $y_2 = x \gg b = B_{2^{m-k+1}-1} || \ldots B_1 || B_0$. Then,

$$B_0 = (x_{2^k - 1} \ x_{2^k - 2} \cdots x_{2^k - a}), \quad a \text{ bit}$$
$$B_{2i-1} = (x_{i*2^k + b - 1} \ x_{i*2^k + b - 2} \cdots x_{i*2^k}), \quad b \text{ bit}$$
$$B_{2i} = (x_{(i+1)*2^k - 1} \ x_{(i+1)*2^k - 2} \cdots x_{(i+1)*2^k - a}), \quad a \text{ bit}$$
$$i \in \{1, \ldots, 2^{m-k} - 1\}$$
$$B_{2^{m-k+1}-1} = (00 \cdots 0), \quad b \text{ bit}$$

For example, if parameters are equal to above case then

$$B_0 = (x_{15} \ x_{14} \cdots x_9),$$
$$B_1 = (x_{24} \ x_{23} \cdots x_{16}),$$
$$B_2 = (x_{31} \ x_{30} \cdots x_{25}),$$
$$B_3 = (000000000).$$

Therefore,

$$y = (x \ll a) \oplus (x \gg b) =$$
$$= y_1 \oplus y_2 = \sum_{i=0}^{2^{m-k+1}-1}(A_i \oplus B_i)$$
$$= A_{2^{m-k+1}-1} || \sum_{i=1}^{2^{m-k+1}-2}(A_i \oplus B_i) || B_0.$$

In $A_0 \oplus B_0$ and $A_{2^{m-k+1}-1} \oplus B_{2^{m-k+1}-1}$, it is obvious that the relations between $x$ and $y$ are 1:1. The only matters are in $A_i \oplus B_i, i \in \{1, \ldots, 2^{m-k+1}-2\}$. It is because of that the equation $u \oplus v = \bar{u} \oplus \bar{v}$ may be held, where $u, v$ is the corresponding bits of $A_i, B_i$ and $\bar{u}, \bar{v}$ are the complements of $u, v$. Namely, one value of $y$ may be correspond to two of $x$ values.

However, as being $A_i = B_{i-2}, i \in \{2, \ldots, 2^{m-k+1}-1\}$, the change of bits in $A_i \oplus B_i$ affect to the bits in $A_{i-2} \oplus B_{i-2} = A_{i-2} \oplus A_i$, $A_0 \oplus B_0$ and $A_{2^{m-k+1}-1} \oplus B_{2^{m-k+1}-1}$, and one value of $y$ can be correspond with only one value of $x$. So, DRT is a 1:1 function.

This is a remarkable property of DRTs. With this property, DRT has the same structure as ROT but exhibits far better dispersion.

## 1.2 DRT has very excellent dispersion.

For ROT, all the bits are only moved parallel within their operands, and the relationship between adjacent bits remains the same as before. DRT has the same structure as ROT, but only $a + b$ bits maintain the former relations and the other bits are all changed maintaining a 1:1 property. Figure 1 and 2 show the excellent dispersion of DRT directly.

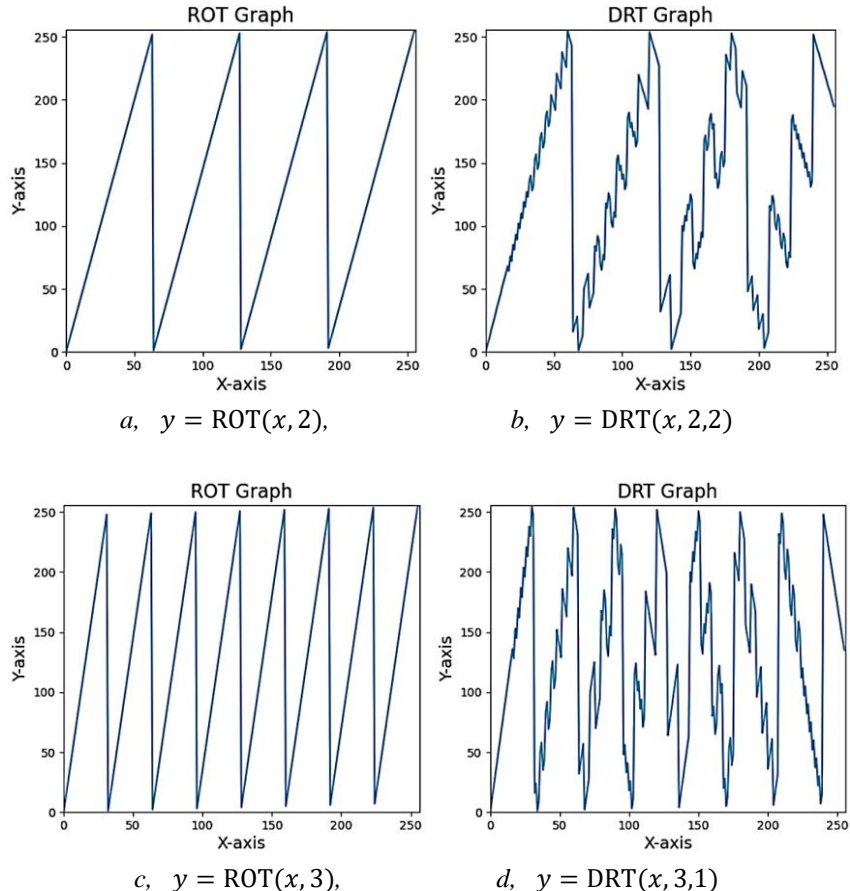

a, $y = \text{ROT}(x, 2)$,     b, $y = \text{DRT}(x, 2, 2)$

c, $y = \text{ROT}(x, 3)$,     d, $y = \text{DRT}(x, 3, 1)$

**Fig. 1** *The graphs of ROT and DRT* ($m = 3, k = 2$)

# 3. Use of the DRT in some stream ciphers

If DRT is used instead of ROT in stream ciphers, the keystream quality can be improved. We have replaced ROT with DRT in HC-128, Rabbit, Salsa20, and compared the qualities of their keystreams with the old ones by using 'National Institute of Standards and Technology (NIST)' randomness testing in [5, 6]. We used the source code submitted to eSTREAM. Key and IV were initialized using the following five methods:

**Method-1:** only one bit in Key[u], IV[v] is set to 1 and others are set to 0. The first byte of Key and IV are set with following 8 pairs and other bytes are all set to 0, so 8 tests are made. For example, Key[u] = 0x01 and IV[v] = 0x10, u, v ∈ {0, 8,15}.

$$\text{Key}[u] \leftarrow \{0x01, 0x02, 0x04, 0x08, 0x10, 0x20, 0x40, 0x80\}$$
$$\text{IV}[v] \leftarrow \{0x10, 0x20, 0x40, 0x80, 0x01, 0x02, 0x04, 0x08\}$$

Next, we make the same as above tests with middle and last bytes. Then 24 tests are made.

**Method-2:** All the bits of Key, IV for each test in method-1 are complemented.

$$\text{Key}[u] \leftarrow \{0xFE, 0xFD, 0xFB, 0xF7, 0xEF, 0xDF, 0xBF, 0x7F\}$$
$$\text{IV}[v] \leftarrow \{0xEF, 0xDF, 0xBF, 0x7F, 0xFE, 0xFD, 0xFB, 0xF7\}$$

**Method-3:** Every bits of Key are set to 0 all the time, and only IVs are set as the same as method-1.

$$\text{Key}[u] \leftarrow \{0x00, 0x00, 0x00, 0x00, 0x00, 0x00, 0x00, 0x00\}$$
$$\text{IV}[v] \leftarrow \{0x10, 0x20, 0x40, 0x80, 0x01, 0x02, 0x04, 0x08\}$$

**Method-4:** Every bits of Key are set to 1 all the time, and only IVs are set as the same as method-2.

$$\text{Key}[u] \leftarrow \{0xFF, 0xFF, 0xFF, 0xFF, 0xFF, 0xFF, 0xFF, 0xFF\}$$
$$\text{IV}[v] \leftarrow \{0xEF, 0xDF, 0xBF, 0x7F, 0xFE, 0xFD, 0xFB, 0xF7\}$$

**Method-5:** Every bytes of Key, IV are set by generator of the random number $Rand()$ function.

Then total 120 tests are made for each cipher and the results of tests are showed as follows.

A/B, B1, B2, B3, hear

    A: the number of tests passed without any fault.

    B: total number of tests having faults.

    B1: the number of tests having only one fault.

    B2: the number of tests having two faults.

    B3: the number of tests having more than two faults.

Then A + B is the number of total test and B = B1 + B2 + B3 in each method.

## 3.1 Using DRT in HC-128.

The rotation parameters 7, 18, 17, and 19 of functions $f_1, f_2$ in the keystream generation algorithm of the HC-128 cipher are replaced with pairs of shift parameters (4, 4), (7, 1), (8, 8), (15,1), and the rotation parameters 23, 10, 8 and 9, 22, 24 of functions $g_1, g_2$ are replaced with pairs of shift parameters (3, 13), (6, 10), and (2, 6), (13, 3), (10, 6), and (6, 2), respectively.

The results of the test for HC-128 using the DRT and ROT by method-1is in table-1. As there are five methods, five such tables can be prepared for HC-128. The same is true for other ciphers. (see Appendix)

**Table 1:** The result of NIST test for HC-128 using DRT and ROT by method-1(128MByte)

| Key[u] | IV[v] | u=0, v=0 | | u=8, v=8 | | u=15, v=15 | |
|---|---|---|---|---|---|---|---|
| | | DRT | ROT | DRT | ROT | DRT | ROT |
| 0x01 | 0x10 | NOT-1 | NOT-1 | REV-1 | 0 | 0 | 0 |
| 0x02 | 0x20 | 0 | NOT-1 RE-1 REV-1 | 0 | 0 | NOT-2 | 0 |
| 0x04 | 0x40 | NOT-1 Uni | 0 | 0 | FFT | 0 | 0 |
| 0x08 | 0x80 | 0 | 0 | 0 | LR NOT-1 | 0 | 0 |
| 0x10 | 0x01 | 0 | FFT NOT-1 | 0 | 0 | RE-1 | 0 |
| 0x20 | 0x02 | 0 | 0 | 0 | 0 | FFT NOT-2 | 0 |
| 0x40 | 0x04 | 0 | 0 | 0 | 0 | 0 | NOT-1 |
| 0x80 | 0x08 | 0 | NOT-1 | 0 | 0 | 0 | NOT-1 |

$\text{Key}[t] = 0x00 \ (t = 0, \ldots, 15, t \neq u), \text{IV}[p] = 0x00 \ (p = 0, \ldots, 15, p \neq v)$

Here 0 mean all items of NIST test are passed, 'XXX' mean item 'XXX' is not passed and 'XXX-y' mean y sub items are not passed. 'XXX' stand for the abbreviation of each item. For example, REV is an abbreviation of the random excursions variant test, and REV-1 indicates that one sub-item is not passed. The following table is an abbreviation for the NIST testing.

**Table 2:** Abbreviation of NIST randomness test names

| No | NIST Test names | Abbreviation |
|---|---|---|
| 1 | The Frequency (Monobit) Test | Freq |
| 2 | Frequency Test within a Block | BF |
| 3 | The Runs Test | Run |
| 4 | Tests for the Longest-Run-of-Ones in a Block | LR |
| 5 | The Binary Matrix Rank Test | Rank |
| 6 | The Discrete Fourier Transform (Spectral) Test | FFT |
| 7 | The Non-overlapping Template Matching Test | NOT |
| 8 | The Overlapping Template Matching Test | OT |
| 9 | Maurer's "Universal Statistical" Test | Uni |
| 10 | The Linear Complexity Test | LC |
| 11 | The Serial Test | Seri |
| 12 | The Approximate Entropy Test | AE |
| 13 | The Cumulative Sums Test | CS |
| 14 | The Random Excursions Test | RE |
| 15 | The Random Excursions Variant Test | REV |

**Table 3:** The result of test for HC-128

| | Using ROT | | | | Using DRT | | | |
|---|---|---|---|---|---|---|---|---|
| | A/B | B1 | B2 | B3 | A/B | B1 | B2 | B3 |
| method-1 | 16/8 | 5 | 2 | 1 | 18/6 | 3 | 3 | 0 |
| method-2 | 16/8 | 4 | 3 | 1 | 14/10 | 7 | 2 | 1 |
| method-3 | 7/17 | 11 | 6 | 0 | 15/9 | 5 | 2 | 2 |
| method-4 | 13/11 | 7 | 2 | 2 | 12/12 | 9 | 2 | 1 |
| method-5 | 16/8 | 6 | 2 | 0 | 13/11 | 9 | 1 | 1 |
| Total | 68/52 | 33 | 15 | 4 | 72/48 | 33 | 10 | 5 |

## 3.2 Using DRT in Rabbit.

In the Rabbit cipher, rotation parameters 16, 16, 8 of $Next\_State()$ function and 16 of $Keysetup()$ function are replaced with pairs of shift parameters (4, 12), (11, 5), (3, 5), and (3, 13). The results of the tests are listed in table 4.

**Table 4:** The result of test for rabbit

|  | Using ROT | | | | Using DRT | | | |
| --- | --- | --- | --- | --- | --- | --- | --- | --- |
|  | A/B | B1 | B2 | B3 | A/B | B1 | B2 | B3 |
| Method-1 | 10/14 | 9 | 3 | 2 | 14/10 | 6 | 4 | 0 |
| Method-2 | 12/12 | 8 | 3 | 1 | 16/8 | 6 | 2 | 0 |
| Method-3 | 12/12 | 7 | 2 | 3 | 12/12 | 10 | 2 | 0 |
| Method-4 | 10/14 | 9 | 1 | 4 | 11/13 | 8 | 2 | 3 |
| Method-5 | 14/10 | 9 | 1 | 0 | 11/13 | 8 | 3 | 2 |
| Total | 58/62 | 42 | 10 | 10 | 64/56 | 38 | 13 | 5 |

## 3.3 Using DRT in Salsa20

In the Salsa20 cipher, rotation parameters 7, 9, 13, and 18 of $quarterround()$ function are replaced with pairs of parameters (4, 4), (6, 2), (10, 6), and (12, 4). The results of the tests are listed in table 5.

**Table 5:** The result of test for Salsa20

|  | Using ROT | | | | Using DRT | | | |
| --- | --- | --- | --- | --- | --- | --- | --- | --- |
|  | A/B | B1 | B2 | B3 | A/B | B1 | B2 | B3 |
| Method-1 | 12/12 | 6 | 5 | 1 | 13/11 | 11 | 0 | 0 |
| Method-2 | 10/14 | 10 | 2 | 2 | 16/8 | 5 | 1 | 2 |
| Method-3 | 9/15 | 12 | 3 | 0 | 12/12 | 8 | 3 | 1 |
| Method-4 | 13/11 | 8 | 3 | 0 | 13/11 | 6 | 3 | 2 |
| Method-5 | 9/15 | 8 | 6 | 1 | 13/11 | 9 | 0 | 2 |
| Total | 53/67 | 44 | 19 | 4 | 67/53 | 39 | 7 | 7 |

As shown in tables 1-5, using DRT instead of ROT improves the quality of the keystreams for all three portfolios. First, more tests were passed without any faults when using DRT instead of ROT. Next, the ratio of the number of tests with a fault to the total number of tests with faults is also increased, and we can estimate that the distribution of faults is also improved. Of course, these results could not have a decisive effect on the safety of stream ciphers, but we can confirm that DRT is more valuable than ROT in terms of the update/output function of stream ciphers.

**Table 6:** The total of tests

|  | A | | B1/B | |
| --- | --- | --- | --- | --- |
|  | Using ROT | Using DRT | Using ROT | Using DRT |
| HC-128 | 68 | 72 | 63% | 69% |
| Rabbit | 58 | 64 | 68% | 68% |
| Salsa20 | 53 | 67 | 66% | 74% |

## 3.4 DRT will improve the safety of HC-128.

In [4], they asserted that the LSB-based distinguishing attack can be expanded to other bits as well. However, using DRT instead of ROT, for $n - a - b$ bits, plus operations are processed along with XOR, so the expansion of the distinguisher of [4] will be made impossible, which will improve the security of HC-128.

# 4. Conclusion

The update or output function is the main component of stream ciphers. The performance of this function is related to its algebraic structures; however, when combined with suitable base operators, better results can be obtained. The DRT is simple, has high dispersion, and would become an attractive base operator for updating the functions of stream ciphers. If one makes proper use of this DRT operator, then very good results would be obtained in stream ciphers.

# Appendix. The detailed results of NIST test

We conducted NIST's coincidence test for a random sequence with a length of 128 Mbytes.

## A.1 The NIST testing for HC-128.

Table A-2 (HC-128, method-2)

| Key[u] | IV[v] | u=0, v=0 | | u=8, v=8 | | u=15, v=15 | |
|---|---|---|---|---|---|---|---|
| | | DRT | ROT | DRT | ROT | DRT | ROT |
| 0xfe | 0xef | 0 | 0 | NOT-1 | 0 | 0 | 0 |
| 0xfd | 0xdf | 0 | 0 | 0 | 0 | NOT-1 | 0 |
| 0xfb | 0xbf | 0 | 0 | RE-1 | 0 | 0 | NOT-1 Uni |
| 0xf7 | 0x7f | 0 | NOT-2 | NOT-2 | NOT-2 | 0 | 0 |
| 0xef | 0xfe | Freq CS-2 | NOT-1 | 0 | 0 | NOT-1 | NOT-1 |
| 0xdf | 0xfd | 0 | OT | NOT-1 | FFT NOT-1 REV-1 | NOT-1 | 0 |
| 0xbf | 0xfb | NOT-1 | 0 | 0 | 0 | 0 | NOT-1 |
| 0x7f | 0xf7 | 0 | 0 | NOT-1 OT | 0 | 0 | 0 |

Table A-3 (HC-128, method-3)

| Key[u] | IV[v] | u=0, v=0 | | u=8, v=8 | | u=15, v=15 | |
|---|---|---|---|---|---|---|---|
| | | DRT | ROT | DRT | ROT | DRT | ROT |
| 0x00 | 0x10 | NOT-3 | NOT-2 | NOT-1 | NOT-2 | 0 | NOT-1 LC |
| 0x00 | 0x20 | NOT-2 REV-5 | RE-1 | 0 | NOT-2 | 0 | 0 |
| 0x00 | 0x40 | 0 | 0 | NOT-1 | NOT-1, RE-1 | 0 | 0 |
| 0x00 | 0x80 | NOT-2 | NOT-1 | NOT-2 | NOT-1 | 0 | 0 |
| 0x00 | 0x01 | 0 | NOT-1 | RE-1 | NOT-1 | 0 | NOT-2 |
| 0x00 | 0x02 | 0 | NOT-1 | 0 | 0 | 0 | NOT-1 |
| 0x00 | 0x04 | NOT-1 | NOT-1 | 0 | 0 | NOT-1 | NOT-1 |
| 0x00 | 0x08 | 0 | NOT-1 | 0 | NOT-1 | 0 | 0 |

IV[p]=0x00 ( p=0,…,15, p≠v )

Table A-4 (HC-128, method-4)

| Key[u] | IV[v] | v=0 | | v=8 | | v=15 | |
|---|---|---|---|---|---|---|---|
| | | DRT | ROT | DRT | ROT | DRT | ROT |
| 0xff | 0xef | 0 | 0 | 0 | 0 | NOT-1 | NOT-1 |
| 0xff | 0xdf | 0 | REV-1 | 0 | NOT-1 | NOT-1 | NOT-2 |
| 0xff | 0xbf | NOT-1 | 0 | 0 | 0 | OT | 0 |
| 0xff | 0x7f | 0 | 0 | 0 | NOT-3 | NOT-3 | 0 |
| 0xff | 0xfe | RE-1 | Run NOT-1 | NOT-2 | OT | 0 | 0 |
| 0xff | 0xfd | 0 | NOT-1 | OT RE-1 | 0 | NOT-1 | NOT-4 |
| 0xff | 0xfb | NOT-1 | 0 | 0 | NOT-1 | 0 | 0 |
| 0xff | 0xf7 | NOT-1 | FFT | 0 | 0 | REV-1 | 0 |

IV[p]=0xff ( p=0,…,15, p≠v )

Table A-5 (HC-128, method-5)

| No | DRT | ROT | No | DRT | ROT | No | DRT | ROT |
|---|---|---|---|---|---|---|---|---|
| 1 | 0 | 0 | 9 | 0 | 0 | 17 | NOT-1 | NOT-1 |

| | | | | | | | |
|---|---|---|---|---|---|---|---|
| 2 | NOT-1 | NOT-1 | 10 | 0 | 0 | 18 | NOT-1 | 0 |
| 3 | NOT-1 | NOT-2 | 11 | NOT-1 | 0 | 19 | 0 | NOT-1 |
| 4 | 0 | REV-1 | 12 | NOT-1 | 0 | 20 | 0 | NOT-2 |
| 5 | 0 | 0 | 13 | NOT-1 | 0 | 21 | 0 | 0 |
| 6 | FFT NOT-2 RE-1 | NOT-1 | 14 | REV-2 | 0 | 22 | 0 | 0 |
| 7 | 0 | 0 | 15 | 0 | 0 | 23 | 0 | 0 |
| 8 | NOT-1 | 0 | 16 | 0 | 0 | 24 | REV-1 | REV-1 |

Number indicate the number of bellow random series.

## A.2 The NIST testing for Rabbit.

Table A-6 (Rabbit, method-1)

| Key[u] | IV[v] | u=0, v=0 | | u=8, v=4 | | u=15, v=7 | |
|---|---|---|---|---|---|---|---|
| | | DRT | ROT | DRT | ROT | DRT | ROT |
| 0x01 | 0x10 | RE-1 | 0 | NOT-1 | 0 | 0 | 0 |
| 0x02 | 0x20 | 0 | NOT-1 | 0 | NOT-1 RE-1 | 0 | 0 |
| 0x04 | 0x40 | 0 | NOT-1 | NOT-1 | 0 | 0 | NOT-1 |
| 0x08 | 0x80 | 0 | 0 | NOT-2 | NOT-2 REV-1 | NOT-1 REV-1 | NOT-1 LC |
| 0x10 | 0x01 | REV-1 | 0 | NOT-2 | Uni | 0 | 0 |
| 0x20 | 0x02 | 0 | 0 | 0 | NOT-1 | 0 | 0 |
| 0x40 | 0x04 | RE-1 | RE-1 | 0 | NOT-2 | 0 | 0 |
| 0x80 | 0x08 | 0 | FFT | NOT-1 | NOT-1 | FFT NOT-1 | 0 |

Key[t] = 0x00 (t = 0,…,15, t ≠ u ), IV[p] = 0x00 (p = 0,…, 7, p ≠ v )

Table A-7 (Rabbit, method-2)

| Key[u] | IV[v] | u=0, v=0 | | u=8, v=4 | | u=15, v=7 | |
|---|---|---|---|---|---|---|---|
| | | DRT | ROT | DRT | ROT | DRT | ROT |
| 0xfe | 0xef | 0 | NOT-2 | NOT-1 | 0 | 0 | 0 |
| 0xfd | 0xdf | 0 | RE-1 | 0 | 0 | 0 | NOT-1 |
| 0xfb | 0xbf | 0 | FFT NOT-1 Uni | RE-1 | NOT-1 | 0 | RE-1 |
| 0xf7 | 0x7f | 0 | 0 | 0 | NOT-1 | NOT-2 | Uni |
| 0xef | 0xfe | NOT-1 | NOT-2 Uni | REV-1 | 0 | 0 | 0 |
| 0xdf | 0xfd | 0 | NOT-2 | RE-1 | 0 | 0 | NOT-1 |
| 0xbf | 0xfb | 0 | Seri | 0 | 0 | 0 | 0 |
| 0x7f | 0xf7 | NOT-1 RE-1 | NOT-1 | 0 | 0 | RE-1 | NOT-1 Seri |

Key[t] = 0xff (t = 0,…, 15, t ≠ u ), IV[p] = 0xff (p = 0,…, 7, p ≠ v )

Table A-8 (Rabbit, method-3)

| Key[u] | IV[v] | v=0 | | v=4 | | v=7 | |
|---|---|---|---|---|---|---|---|
| | | DRT | ROT | DRT | ROT | DRT | ROT |
| 0x00 | 0x10 | NOT-1 | 0 | 0 | NOT-1 | NOT-1 Uni | 0 |
| 0x00 | 0x20 | 0 | 0 | NOT-1 | 0 | 0 | NOT-3 RE-2 |
| 0x00 | 0x40 | NOT-1 | 0 | 0 | 0 | 0 | 0 |
| 0x00 | 0x80 | NOT-1 | NOT-1 | OT | NOT-1 | 0 | 0 |
| 0x00 | 0x01 | 0 | NOT-2 | 0 | 0 | FFT | 0 |
| 0x00 | 0x02 | NOT-1 | NOT-1 | RE-1 | RE-1 | 0 | RE-1 |
| 0x00 | 0x04 | NOT-2 | NOT-2 RE-1 | REV-1 | 0 | NOT-1 | NOT-3 |

| | | | | | | NOT-1 | | NOT-2 |
|---|---|---|---|---|---|---|---|---|
| 0x00 | 0x08 | 0 | 0 | 0 | NOT-1 | 0 | NOT-2 |

IV[p]=0x00 ( p=0,…,7, p ≠ v )

Table A-9 (Rabbit, method-4)

| Key[u] | IV[v] | v=0 | | v=4 | | v=7 | |
|---|---|---|---|---|---|---|---|
| | | DRT | ROT | DRT | ROT | DRT | ROT |
| 0xff | 0xef | 0 | NOT-1 | 0 | 0 | Uni | 0 |
| 0xff | 0xdf | RE-1 | NOT-1 | NOT-1 | NOT-1 | 0 | 0 |
| 0xff | 0xbf | Run | 0 | 0 | 0 | NOT-4 | NOT-2 |
| 0xff | 0x7f | 0 | NOT-2 REV-1 | 0 | 0 | NOT-3 REV-1 | RE-1 |
| 0xff | 0xfe | NOT-1 | NOT-4 | Frq CS-2 NOT-1 | CS NOT-1 REV-2 | NOT-1 RE-1 | NOT-1 |
| 0xff | 0xfd | 0 | 0 | NOT-1 RE-1 | 0 | 0 | NOT-2 OT |
| 0xff | 0xfb | RE-1 | 0 | 0 | NOT-1 | 0 | NOT-1 |
| 0xff | 0xf7 | 0 | 0 | NOT-1 | RE-1 | NOT-1 | NOT-1 |

IV[p] = 0xff ( p = 0,…, 7, p ≠ v )

Table A-10 (Rabbit, method-5)

| No | DRT | ROT | No | DRT | ROT | No | DRT | ROT |
|---|---|---|---|---|---|---|---|---|
| 1 | NOT-1 RE-1 | OT | 9 | 0 | 0 | 17 | NOT-1 | 0 |
| 2 | 0 | LR | 10 | 0 | 0 | 18 | NOT-1 RE-1 | 0 |
| 3 | REV-1 | 0 | 11 | 0 | NOT-1 | 19 | 0 | 0 |
| 4 | NOT-2 RE-1 | 0 | 12 | 0 | NOT-1 | 20 | NOT-3 | 0 |
| 5 | REV-1 | NOT-1 | 13 | NOT-1 | 0 | 21 | 0 | 0 |
| 6 | 0 | 0 | 14 | NOT-1 | NOT-2 | 22 | NOT-1 | NOT-1 |
| 7 | NOT-1 REV-1 | FFT, | 15 | NOT-1 | 0 | 23 | 0 | 0 |
| 8 | 0 | OT | 16 | 0 | 0 | 24 | Uni | NOT-1 |

Number indicate the number of bellow random series.

## A.3 The NIST testing for Salsa20

Table A-11 (Salsa20, method-1)

| Key[u] | IV[v] | u=0, v=0 | | u=8, v=4 | | u=15, v=7 | |
|---|---|---|---|---|---|---|---|
| | | DRT | ROT | DRT | ROT | DRT | ROT |
| 0x01 | 0x10 | 0 | 0 | NOT-1 | 0 | 0 | 0 |
| 0x02 | 0x20 | 0 | NOT-1 | 0 | NOT-1 | NOT-1 | 0 |
| 0x04 | 0x40 | 0 | NOT-2 | RE-1 | 0 | 0 | 0 |
| 0x08 | 0x80 | 0 | NOT-2 | 0 | 0 | RE-1 | 0 |
| 0x10 | 0x01 | 0 | NOT-1 | 0 | 0 | NOT-1 | CS-1 NOT-1 |
| 0x20 | 0x02 | NOT-1 | RE-1 REV-1 Seri | 0 | NOT-1 | RE-1 | Seri |
| 0x40 | 0x04 | NOT-1 | 0 | Frq | NOT-1 | NOT-1 | NOT-1 OT |
| 0x80 | 0x08 | 0 | RE-1 NOT-1 | NOT-1 | 0 | 0 | 0 |

Key[t] = 0x00 (t = 0,…, 15, t ≠ u ), IV[p] = 0x00 (p = 0,…, 7, p ≠ v )

Table A-12 (Salsa20, method-2)

| Key[u] | IV[v] | u=0, v=0 | | u=8, v=4 | | u=15, v=7 | |
|---|---|---|---|---|---|---|---|
| | | DRT | ROT | DRT | ROT | DRT | ROT |
| 0xfe | 0xef | 0 | NOT-1 | 0 | NOT-1 | 0 | 0 |
| 0xfd | 0xdf | 0 | Frq CS-1 NOT-1 | NOT-1 | 0 | NOT-1 | 0 |
| 0xfb | 0xbf | 0 | RE-1 | NOT-2 | 0 | 0 | 0 |
| 0xf7 | 0x7f | 0 | NOT-1 | 0 | NOT-1 | NOT-2 RE-1 | 0 |
| 0xef | 0xfe | 0 | 0 | 0 | NOT-1 | NOT-2 RE-1 | NOT-1 |
| 0xdf | 0xfd | 0 | 0 | 0 | NOT-1 | 0 | 0 |
| 0xbf | 0xfb | NOT-1 | NOT-1 OT | 0 | OT | 0 | 0 |
| 0x7f | 0xf7 | REV-1 | LR NOT-3 Uni | NOT-1 | RE-1 | 0 | RE-1 Seri |

Key[t] = 0xff (t = 0,..., 15, t ≠ u ), IV[p] = 0xff (p = 0,..., 7, p ≠ v )

Table A-13 (Salsa20, method-3)

| Key[u] | IV[v] | v=0 | | v=4 | | v=7 | |
|---|---|---|---|---|---|---|---|
| | | DRT | ROT | DRT | ROT | DRT | ROT |
| 0x00 | 0x10 | 0 | NOT-2 | NOT-2 RE-1 | NOT-1 | FFT | CS-1 |
| 0x00 | 0x20 | NOT-1 | NOT-1 REV-1 | NOT-1 | NOT-1 | 0 | 0 |
| 0x00 | 0x40 | 0 | NOT-2 | 0 | 0 | 0 | NOT-1 |
| 0x00 | 0x80 | NOT-2 | RE-1 | 0 | NOT-1 | REV-1 | 0 |
| 0x00 | 0x01 | 0 | NOT-1 | REV-2 | 0 | 0 | 0 |
| 0x00 | 0x02 | 0 | Rank | RE-1 | 0 | NOT-1 | NOT-2 |
| 0x00 | 0x04 | 0 | NOT-1 | NOT-2 | 0 | 0 | 0 |
| 0x00 | 0x08 | 0 | NOT-1 | NOT-1 | NOT-1 | NOT-1 | 0 |

IV[p] = 0x00 ( p = 0,..., 7, p ≠ v )

Table A-14 (Salsa20, method-4)

| Key[u] | IV[v] | v=0 | | v=4 | | v=7 | |
|---|---|---|---|---|---|---|---|
| | | DRT | ROT | DRT | ROT | DRT | ROT |
| 0xff | 0xef | NOT-1 REV-1 | 0 | 0 | OT | 0 | 0 |
| 0xff | 0xdf | NOT-1 | 0 | 0 | AE | 0 | 0 |
| 0xff | 0xbf | RE-1 | REV-1 | NOT-2 | 0 | NOT-1 | 0 |
| 0xff | 0x7f | 0 | 0 | RE-1 | 0 | FFT NOT-2 REV-1 | NOT-1 |
| 0xff | 0xfe | 0 | NOT-1 REV-1 | NOT-1 | 0 | 0 | 0 |
| 0xff | 0xfd | NOT-1 | 0 | NOT-2 | NOT-1 | 0 | 0 |
| 0xff | 0xfb | 0 | NOT-1 | 0 | RE-1 REV-1 | 0 | NOT-2 |
| 0xff | 0xf7 | 0 | NOT-1 | 0 | 0 | NOT-3 | NOT-1 |

IV[p] = 0xff ( p = 0,..., 7, p ≠ v )

Table A-15 (Salsa20, method-5)

| No | DRT | ROT | No | DRT | ROT | No | DRT | ROT |
|---|---|---|---|---|---|---|---|---|
| 1 | 0 | 0 | 9 | 0 | NOT-1 OT RE-1 | 17 | 0 | 0 |
| 2 | 0 | 0 | 10 | 0 | NOT-1 | 18 | 0 | NOT-1 |

| 3 | 0 | 0 | 11 | 0 | 0 | 19 | RE-1 | NOT-2 |
|---|---|---|---|---|---|---|---|---|
| 4 | NOT-2 REV-1 | 0 | 12 | 0 | NOT-2 | 20 | 0 | NOT-2 |
| 5 | NOT-1 | FFT | 13 | 0 | NOT-2 | 21 | REV-1 | NOT-1 |
| 6 | 0 | 0 | 14 | NOT-1 | Uni | 22 | NOT-1 | NOT-1 REV-1 |
| 7 | 0 | 0 | 15 | NOT-1 | Run | 23 | Rank NOT-2 | NOT-1 |
| 8 | NOT-1 | NOT-1 | 16 | NOT-1 | REV-2 | 24 | NOT-1 | 0 |

Number indicate the number of bellow random series.

## A.4 The random series used in method-5.

In the method-5, the following random series are used. The first is the number of series and the next two lines are 16 bytes of hexadecimal numbers for key and iv individually.

1
Key  F2 A7 96 D2 7A 1C 53 87 D3 E2 CE EE 54 86 B0 1E
iv  7D 14 25 A4 81 8A 82 8E 01 8D 2A C3 1B 5B B2 96
2
Key  56 20 D0 87 3B 0A 8C 26 26 27 D5 AF A5 A4 B2 5A
iv  17 32 A1 02 03 09 8A 3D AE 2C F2 CD EC EF 71 37
3
Key  93 C3 A9 44 7C E7 E0 FD 7F 9C CA 5F 34 25 72 DD
iv  98 A0 4D A7 6B DC C0 1D D6 6A E3 8A 51 F7 85 0F
4
Key  8A 9A 13 DB 59 13 35 0C 40 2A D0 DE 1F CB 6B CF
iv  4B 8D 98 36 8F 9C 1B D5 70 F6 4F B1 B5 DB C6 22
5
Key  96 73 31 A2 3C DB 2A 68 1E DC 70 FA 7B CA D8 6B
iv  6E 61 FF 75 16 59 8B B2 F4 3A B4 0F 94 72 04 7F
6
Key  D2 17 0A 60 7E B8 7E 3E 76 52 65 AB 0A 4C 98 ED
iv  F0 D0 AB 1A 7E 2C C1 92 1C 78 A5 CD FA 79 17 57
7
Key  7B 43 B0 07 5C C3 08 BC 41 41 49 09 70 BA 0D 6D
iv  73 5F 54 39 6E 96 79 AC AF 42 65 5C 87 44 02 D7
8
Key  A4 FC D8 C8 DF 97 69 2E 9C CF B6 B1 1D 6E AC 94
iv  68 F6 18 82 49 93 C6 24 96 4F EA F3 B6 85 C0 4B
9
Key  6D B9 10 C7 1C A3 AB 2E 82 D9 0D 37 1C 36 B9 2C
iv  94 09 CF 79 37 BB 26 6E A9 B2 9C 41 D7 74 AA 52
10
Key  0B F0 5E 70 1B AA BC FA CE D5 AD 10 11 BE 1D 7D
iv  D1 AD 05 EA 60 BA E9 54 1B E3 A6 BD 4A 5C 6A 21
11
Key  5E 62 AD F3 1F 52 7E DE 83 F0 86 61 6C 25 5B CA
iv  BC DB 8C 7C 16 B5 84 45 E8 FD B1 EB A9 DE E6 08
12
Key  3F 6A A1 C3 39 B3 65 E0 E7 D8 A8 40 8B E5 D6 F2
iv  05 08 FB 1E EC 4E 7B EC DD AD 03 A1 13 47 BF 0C

13
Key  2D 18 76 F0 F2 0C 10 00 69 F3 B5 3F 60 32 C8 D6
iv  6B E1 91 3D 33 CE 32 5C 11 C5 DC F0 DC 0F 42 31
14
Key  91 90 B1 A5 B3 F9 49 9F BC 37 BC 00 B2 4F CA 12
iv  04 FE 0D 9A B6 4D 3A 0B BD 64 A4 FA AD A3 01 D2
15
Key  A3 62 5F 6C 2B 6E 8F 1C 9A E2 33 74 03 93 97 21
iv  AC C5 C2 A2 6C FB F7 B5 57 0E 82 0F 9E 7D B1 50
16
Key  81 6E 8B 3D FA 78 4E 8D 7E CD EA 27 A7 B1 63 90
iv  33 F8 0A 60 AA 1B 48 F6 41 8B 97 6A FF 45 D1 C3
17
Key  E5 E6 C5 F2 BB 66 87 2C D1 12 F0 E8 F8 CE 64 CB
iv  CC 15 87 BD 2C 99 50 A5 EE 2A 5F 73 D0 D9 90 64
18
Key  08 A6 5D B5 A8 8D 97 7D 2D A8 84 38 B0 3E A3 7E
iv  3E BA FD 3D D8 A5 4F 50 BE D2 6B 53 EE D8 DC B6
19
Key  5C FB 78 E8 12 95 B6 42 01 0D C2 05 45 DC BC 1E
iv  21 11 0B AD 4B 51 4D 1E 8A 1C 94 95 83 51 CF 90
20
Key  AA 58 FC 36 85 54 A3 AE 17 EF 1F 95 96 17 F6 64
iv  65 36 37 06 93 AC 16 EB A5 7C D6 58 95 45 EC 01
21
Key  A1 2F 67 CC 61 81 F9 BE D8 7D 25 14 81 BD EF 56
iv  18 23 82 96 B7 6C 71 A3 3F 08 42 7F F9 29 2D 13
22
Key  9C 1A 9C 98 CF 97 23 46 38 44 A8 54 77 90 EC 4F
iv  71 19 27 5D 49 CD 9F 7F 8C 4E 78 13 AB 9B CE 9D
23
Key  91 0C 41 64 A8 00 FD AC 99 13 53 3B 76 1E 88 D4
iv  47 1D 7F 5B AB 3B 7F 8D C3 2F 35 EF 4B CB FD 04
24
Key  90 F4 3E 2F 61 6D 50 C5 79 D5 43 A7 E7 94 35 87
iv  62 0C 4B 08 D5 14 53 D0 1B A8 A8 DE 06 DE 56 1E